\documentclass[twocolumn,aps,pre,showpacs,floatfix]{revtex4}
\usepackage{graphicx}
\usepackage{amssymb}
\usepackage{amsmath}
\usepackage{bm}

\begin{document}

\title{Linear stability in networks of pulse-coupled neurons}


\author{Simona Olmi}
\email{simona.olmi@fi.isc.cnr.it}    
\affiliation{CNR - Consiglio Nazionale delle Ricerche -
Istituto dei Sistemi Complessi, via Madonna del Piano 10,
I-50019 Sesto Fiorentino, Italy}
\affiliation{INFN - Sezione di Firenze and CSDC, via Sansone 1, 50019 Sesto      Fiorentino, Italy}

\author{Antonio Politi}
\email{a.politi@abdn.ac.uk}    
\affiliation{SUPA and Institute for Complex Systems and Mathematical Biology, King's College,
University of Aberdeen, Aberdeen AB24 3UE, United Kingdom}
\affiliation{CNR - Consiglio Nazionale delle Ricerche -
Istituto dei Sistemi Complessi, via Madonna del Piano 10,
I-50019 Sesto Fiorentino, Italy}

\author{Alessandro Torcini}
\email{alessandro.torcini@cnr.it}    
\affiliation{CNR - Consiglio Nazionale delle Ricerche -
Istituto dei Sistemi Complessi, via Madonna del Piano 10,
I-50019 Sesto Fiorentino, Italy}
\affiliation{INFN - Sezione di Firenze and CSDC, via Sansone 1, 50019 Sesto      Fiorentino, Italy}

\begin{abstract}
In a first step towards the comprehension of neural activity, one should focus on
the stability of the various dynamical states. Even the characterization of idealized regimes, 
such as a perfectly periodic spiking activity, reveals unexpected difficulties. In this paper we 
discuss a general approach to linear stability of pulse-coupled neural networks for generic 
phase-response curves and post-synaptic response functions. In particular, we present: (i)
a mean-field approach developed under the hypothesis of an infinite network and small 
synaptic conductances; (ii) a ``microscopic" approach which applies to finite but large networks.
As a result, we find that no matter how large is a neural network, its response to most of the 
perturbations depends on the system size. There exists, however, also a second class of perturbations,
whose evolution typically covers an increasingly wide range of time scales.
The analysis of perfectly regular, asynchronous, states reveals that their stability
depends crucially on the smoothness of both the phase-response curve and the transmitted 
post-synaptic pulse. The general validity of this scenarion is confirmed by numerical simulations 
of systems that are not amenable to a perturbative approach.
\end{abstract}




\maketitle

\section{Introduction}
\label{intro}

Networks of oscillators play an important role in both biological (neural systems, circadian rythms,
population dynamics) ~\cite{pikov} and physical contexts (power grids, Josephson junctions, cold atoms)
\cite{filatrella,hadley,politi_cold}.
It is therefore comprehensible that many studies have been and are still devoted to 
understanding their dynamical properties. Since the development of sufficiently powerful
tools and the resulting discovery of general laws is an utterly difficult task, it is
convenient to start from simple setups.

The first issue to consider is the oscillator model-structure. As phases
are typically more sensitive than amplitudes to mutual coupling, they are
likely to provide the most relevant contribution to the collective evolution~\cite{pikov}.
Accordingly, here we restrict our analysis to oscillators characterized by a single,
phase-like, variable. This is tipically done by reducing the neuronal dynamics to the evolution
of the membrane potential and introducing the corresponding {\it velocity field}
which describes the single-neuron activity. Equivalently, one can map the membrane potential
onto a phase variable and simultaneously introduce a phase-response curve (PRC)~\cite{note1} to take 
into account the dependence of the neuronal response on the current value of the membrane potential
(i.e. the phase). In this paper 
we adopt the first point of view, with a few exceptions, when the second one is mathematically
more convenient.

As for the coupling, two mechanisms are typically invoked in the literature, diffusive and pulse-mediated. 
While the former mechanism is pretty well understood (see e.g. the very many papers devoted to 
Kuramoto-like models \cite{acebron}), the latter one, more appropriate in neural dynamics, 
involves a series of subtleties that have not yet been fully appreciated. This is why here we 
concentrate on pulse-coupled oscillators.

Finally, for what concerns the topology of the interactions, globally coupled identical oscillators 
provide a much simplified but already challenging test bed. The high symmetry of the 
corresponding evolution equations simplifies the identification of the stationary solutions and the 
analysis of their stability properties. The two most symmetric solutions are: (i) the fully synchronous state, 
where all oscillators follow exactly the same trajectory; (ii) the splay state (also known as 
``ponies on a merry-go-round", antiphase state or rotating waves)~\cite{krupa,hadley,ashw90},
where the oscillators still follow the same periodic trajectory, 
but with different (evenly distributed) time shifts.
The former solution is the simplest representative of the broad class of clustered states \cite{golomb}, 
where several oscillators behave in the same way, while the latter is the prototype of asynchronous states, 
characterized by a smooth distribution of phases \cite{renart}.

In spite of the many restrictions on the mathematical setup, the stability of the synchronous and splay 
states still depend significantly on additional features such as the synaptic reponse-function,
the velocity field, and the presence of delay in the pulse transmission. As a result, one can encounter 
splay states that are either strongly stable along all directions, or that present many almost-marginal 
directions, or, finally, that are marginally stable along various directions~\cite{nicols, watanabe}. 
Several analytic results have been obtained in specific cases, but a global picture is still missing:
the goal of this paper is to recompose the puzzle, by exploring the role of the velocity field (or,
equivalently, of the phase response curve) and of the shape of the transmitted post-synaptic potentials.

In pulse-coupled oscillators, even the stability assessment of the fully synchronous regime  
is far from trivial: in fact, the pulse emission introduces a discontinuity which requires separating
the evolution before and after such event. Moreover, when many neurons spike at the same time, the 
length of some interspike intervals is virtually zero but cannot be neglected in the 
mathematical analysis. In fact, the first study of this problem was restricted to excitatory coupling 
and $\delta$-pulses \cite{mirollo}. In that context, the stability of the synchronous state
follows from the fact that when the phases of two oscillators are sufficiently close to one
another, they are instantaneously reset to the same value (as a result of a non-physical lack of
invertibility of the dynamics). 
The first, truly linear stability analyses have been performed later, first in the case of
two oscillators \cite{vres94,hansel95} and then considering $\delta$-pulses with continuous PRCs~\cite{goel}. 
Here, we extend the analysis to generic pulse-shapes and discontinuous PRCs 
(such as for leaky integrate and fire (LIF) neurons).

As for the splay states, their stability can be assessed in two ways: (i) by assuming that the number of
oscillators is infinite (i.e. taking the so called thermodynamic limit) and thereby studying the evolution of 
the distribution of the membrane potentials -- this approach is somehow equivalent to dealing with (macroscopic)
Liouville-type equations in statistical mechanics; (ii) by dealing with the (microscopic) equations of
motion for a large but finite number $N$ of oscillators.
As shown in some pioneering works \cite{kura1991,treves}, the former approach corresponds to develop a
mean field theory. The resulting equations have been first solved in~\cite{abbott} for
pulses composed of two exponential functions, in the limit of a small effective coupling~\cite{note2}.
Here, following~\cite{abbott}, we extend the analysis to generic pulse-shapes, finding that
substantial differences exist among $\delta$, exponential and the so-called $\alpha$-pulses (see the next
section for a proper definition).

Direct numerical studies of the linear stability of finite networks suggest that the eigenfunctions 
of the (Floquet) operator can be classified according to their wavelength $\ell$ (where $\ell$ refers 
to the neuronal phase - see Sec.~\ref{sec:3:A} for a precise definition). In finite systems, it is 
convenient to distinguish between long (LW) and short (SW) wavelengths. Upon considering that $\ell = n/N$ 
($1\le n\le N)$, LW can be identified as those for which $n\ll N$, while SW correspond to larger $n$ values.
Numerical simulations suggest also that the time scale of a LW perturbation typically increases upon 
increasing its wavelength, starting from a few milliseconds (for small $n$ values) up to much 
longer values (when $n$ is on the order of the network size $N$) which depend on ``details" 
such as the continuity of the velocity field, or the pulse shape. On the other hand, SW are 
characterized by a slow size-dependent dynamics. 

For instance, in LIF neurons coupled via $\alpha$-pulses, it has been found \cite{calamai} that
the Floquet exponents of LW decrease as $1/\ell^2$ (for large $\ell$), while the time scale of
the SW component is on the order of $N^2$. In practice the LW spectral component as determined
from the finite $N$ analysis coincides with that one obtained with the mean field approach 
(i.e. taking first the thermodynamic limit). As for the SW component, it cannot be quantitatively
determined by the mean-field approach, but it is nevertheless possible to infer the correct order
of magnitude of this time scale. In fact, upon combining the $1/\ell^2$ decay (predicted by the
mean-field approach) with the observation that the minimal wavelength is $1/N$, it naturally follows
that the SW time scale is $N^2$, as analytically proved in \cite{olmi_math}. Furthermore, it has
been found that the the two spectral components smoothly connect to each other and the predictions
of the two theoretical approaches coincide in the crossover region.

It is therefore important to investigate whether the same agreement extends to more generic pulse shapes
and velocity fields. The finite-$N$ approach can, in principle, be generalized to arbitrary
shapes, but the analytic calculations would be quite lengthy, due to the need of distinguishing
between fast and slow scales and the need of accounting for higher order terms. 
For this reason, here we give a positive answer to this question only with the help of numerical
studies. 

The only, important, exception to this scenario is obtained for quasi $\delta$-like pulses \cite{zillmer2}, 
i.e. for pulses whose width is smaller than the average time sepration between any two consecutive spikes,
in which case all the SW eigenvalues remain finite for increasing $N$. 

In Sec.~\ref{sec:1} we introduce the model and derive the corresponding event-driven map, a necessary
step before undertaking the analytic calculations. Sec.~\ref{sec:2} is devoted to a perturbative stability 
analysis of the splay state in the infinite-size limit for generic velocity fields and pulse shapes.
The following Sec.~\ref{sec:3} reports a discussion of the stability in finite networks.
There we briefly recall the main results obtained in \cite{olmi_math} for the splay state and we extensively
discuss the method to quantify the stability of the fully synchronous regime.
The following two sections are devoted to a numerical analysis of various setups. In Sec.~\ref{sec:4}
we study splay states in finite networks for generic velocity fields and three different classes of
of pulses, namely, with finite, vanishing ($\approx 1/N$), and zero width. In Sec.~\ref{sec:5} we study
periodically forced networks. Such studies show that the scaling relations derived for the
splay states apply also to such a microscopically quasi-periodic regime. A brief summary of the main
results together with a recapitulation of the open problem is finally presented in Sec.~\ref{sec:6}.
In the first appendix we derive the Fourier components needed to assess the stability of a splay state
for a generic PRC. In the second appendix the evaporation exponent is determined for the synchronous state
in LIF neurons. 

\section{The Model}
\label{sec:1}

The general setup considered in this paper is a network of $N$ identical pulse-coupled neurons 
(rotators), whose evolution is described by the equation 
\begin{equation}
\label{sistema}
\dot{X}^j = F(X^j) + gE(t), \qquad j=1,\ldots,N
\end{equation}
where $X^j$ represents the membrane potential, $g$ is the coupling constant and $E(t)$ is the 
{\it mean field}. When $X^j$ reaches the threshold value $X^j=1$, it is reset to $X^j=0$ 
and a spike contributes to the mean field $E$ in a way that is described here below. 
The resetting procedure is an approximation of the discharge mechanism operating in real neurons. 
The function $F(X)$ (the velocity field) is assumed to be everywhere positive, thus ensuring that the neuron 
is repetitively firing.  For $F_0(X)=a-X$ the model reduces to the well-known case of LIF neurons. 

The mean field $E$ arises from the linear superposition of the pulses
emitted by the single neurons. In full generality, we assume that
\begin{equation}
E^{(L)} = \sum_i^{L-1} a_i E^{(i)} + \frac{K}{N}\sum_{n|t_n<t} \delta(t-t_n) \ ,
\label{dem0}
\end{equation}
where the superscript $(i)$ denotes the $i$th time derivative, and $L$ the order of the differential equation.
Moreover, $K =\prod_i \alpha_i$, ($-\alpha_i$ being the poles of the differential equation), so as to 
ensure that the single pulse has unit area (for $N=1$) while $t_n$ denotes the time at which the $n$th spike 
is emitted. $L$ controls the smoothness of the pulses: $L-1$ is the order of the lowest 
derivative that is discontinuous: $L=0$ corresponds to the extreme case of $\delta$-pulses 
with no field dynamics; $L=1$ corresponds to discontinuous exponential pulses; 
$L=2$ (with $\alpha_1=\alpha_2$) to the so-called $\alpha$-pulses ($E_s(t)=\alpha^2 t e^{-\alpha t}$).
Since $\alpha$-pulses will be often referred to, it is worth being a little more specific. 
In this case, Eq.~(\ref{dem0}) reduces to
\begin{equation}
\label{eq:E}
  \ddot E(t) +2\alpha\dot E(t)+\alpha^2 E(t)= 
  \frac{\alpha^2}{N}\sum_{n|t_n<t} \delta(t-t_n) \ ,
\end{equation}
and it is convenient to transform this equation into a system of two first order differential 
equations, namely
\begin{equation}
 \label{sistema2}
\dot{E}= P -\alpha E, \qquad \dot{P}+\alpha P = \frac{\alpha^2}{N}\sum_{n|t_n<t} \delta(t-t_n) \ ,
\end{equation}
where we have introduced, for the sake of simplicity, the new variable $P\equiv \alpha E+ \dot{E}$.

\subsection{Event-driven Map}
By following Ref. \cite{calamai, zillmer}, it is convenient to pass from a continuous to a discrete
time evolution rule, by deriving the event-driven map which connects the network
configuration at consecutive spike times. For the sake of simplicity, in the following part of
this section we refer to $\alpha$-pulses, but there is no conceptual limitation to extending
the approach to $L>2$.

By integrating Eq.~(\ref{sistema2}), we obtain
\begin{eqnarray}
\label{event_driven_map_E}
  &&E_{n+1}=E_n {\rm e}^{-\alpha\mathcal{T}_n }+P_n\mathcal{T}_n 
  {\rm e}^{-\alpha \mathcal{T}_n} 
  \\ \label{event_driven_map_P}
  &&P_{n+1}=P_n e^{-\alpha \mathcal{T}_n}+ \frac{\alpha^2}{N}  \, ,
\end{eqnarray}
where we have taken into account the effect of the incoming pulse
(see the term $\alpha^2/N$ in the second equation) while
$\mathcal{T}_n= t_{n+1}-t_n$ is the interspike interval; $t_{n+1}$
corresponds to the time when the neuron with the largest membrane
potential reaches the threshold.

Since all neurons follow the same first-order differential equation (this is a mean-field
model), the ordering of their membrane potentials is preserved (neurons ``rotate'' around the 
circle $[0,1]$ without overtaking each other~\cite{zin}). It is, therefore, convenient to 
order the potentials from the largest to the smallest one and to introduce a co-moving reference 
frame, i.e. to shift backward the label $j$, each time a neuron reaches the threshold. By formally
integrating Eq.~(\ref{sistema}), 
\begin{eqnarray}
\nonumber
 X^j_{n+1}&=&\mathcal{F}(X^{j+1}_n,\mathcal{T}_n) +g \frac{{\rm e}^{-\mathcal{T}_n} 
 - e^{-\alpha\mathcal{T}_n}}{\alpha-1} \left(E_n+\frac{P_n}{\alpha-1} \right) 
\\   \label{generalized_event_driven_map}    
 &-& g \frac{\mathcal{T}_n e^{-\alpha\mathcal{T}_n}}{(\alpha-1)} P_n \, .
\end{eqnarray}
Moreover, since $X^1_n$ is always the largest potential, the interspike interval is 
defined by the threshold condition
\begin{equation}
\label{tauu}
X^1_n(\mathcal{T}_n,E_n,P_n) \equiv 1 
\qquad .
\end{equation}
Altogether, the model now reads as a discrete-time map, involving $N+1$ variables:
$E_n$, $P_n$, $X^j_n$ ($1\le j<N$), as one degree of freedom has been eliminated as a result of having 
taken the Poincar\'{e} section, since $X^N_n \equiv 0$ due to the resetting mechanism. The advantage of 
the map description is that we do not have to deal
any longer with $\delta$-like discontinuities, or with formally infinite sequences of past events.

In this framework the splay state is a fixed point of the event-driven map.
Its coordinates can be determined in the following way. From
Eq.~(\ref{event_driven_map_E}), one can express $\tilde P$ and $\tilde E$ as a function of
the yet unknown interspike interval $\mathcal{T}$,
\begin{equation}
\label{field_splay}
\tilde P = \frac{\alpha^2}{N} (1-{\rm e}^{-\alpha \mathcal{T}})^{-1}
\quad \tilde E = \mathcal{T} {\tilde P} ({\rm e}^{\alpha \mathcal{T}}-1)^{-1}
\enskip .
\end{equation}
The value of the membrane potentials $\tilde X^k$ are then obtained by iterating backward in $j$ 
Eq.~(\ref{generalized_event_driven_map}) (the $n$ dependence is dropped for the fixed point)
starting from the initial condition $\tilde X^N=0$. The interspike interval $\mathcal{T}$ is 
finally obtained by imposing the condition $\tilde X^0=0$.
In practice the computational difficulty amounts to finding the zero of a one dimensional
function and, even though  $\mathcal{F}(X^{j+1},\mathcal{T})$ can, in most cases, be
obtained only through numerical integration, the final error can be very well kept under
control.

\section{Theory ($N=\infty$)}
\label{sec:2}

The stability of a dynamical state can be assessed by either first taking the infinite
time limit and then the thermodynamic limit, or vice versa. In general it is not obvious whether
the two methods yield the same result and this is particularly crucial for the splay state, as many 
eigenvalues tend to 0 for $N\to\infty$. In this section we discuss the scenarios that have to
be expected when the thermodynamic limit is taken first. We do that, by following Abbott 
and van Vreeswijk~\cite{abbott}. 

As a first step, it is convenient to introduce the phase-like variable
\begin{equation}
y^i = \int_0^{X^i} \frac{dx}{G(x)} \quad, \quad 0\le y^i\le 1
\end{equation}
where, for later convenience, we have defined $G(X) \equiv g + T_0 F(X)$, 
$T_0=N\mathcal{T}$ being the 
period of the splay state (i.e. the single-neuron interspike interval).
The phase $y^i$ evolves according to the equation
\begin{equation}
\frac{d y^i}{dt} = \tilde E + \frac{g\varepsilon(t)}{G(X(y^i))}
\label{phase}
\end{equation}
where $\tilde E = 1/T_0$ is the amplitude of the field in the splay state,
$\varepsilon(t) = E(t) - \tilde E$.  In the splay state, since $\varepsilon=0$, 
$y^i$ grows linearly in time, as indeed expected for a well-defined phase.
In the thermodynamic limit, the evolution is ruled by the continuity equation
\begin{equation}
\frac{\partial \rho}{\partial t} = -\frac{\partial J}{\partial y}
\label{eq:cont}
\end{equation}
where $\rho(y,t) dy$ is the fraction of neurons whose phase $y^i$ lies in 
$(y, y + dy)$ at time $t$, and
\begin{equation}
J(y,t) = \left[ \tilde E + \frac{g \varepsilon(t)}{G(X(y))} \right] \rho(y,t) 
\label{flux}
\end{equation}
is the corresponding flux. As the resetting implies that the outgoing flux 
$J(1,t)$  (which coincides with the firing rate) equals the incoming
flux at the origin, the above equation has to be complemented with the boundary condition 
$J(0,t)=J(1,t)$. Finally, in this macroscopic representation, the field equation writes 
\begin{equation}
\varepsilon^{(L)} = \sum_i^{L-1} a_i \varepsilon^{(i)} + K (J(1,t)-\tilde E) \ ,
\label{dem}
\end{equation}
while the splay state corresponds to the fixed point $\rho = 1$, $\varepsilon=0$, $J = \tilde E$.
Its stability can be studied by introducing the perturbation $j(y,t)$
\begin{equation}
j(y,t) = J(y,t) - \tilde E \, ,
\end{equation}
and linearizing the continuity equation, 
\begin{equation}
\frac{\partial j}{\partial t} = \frac{g}{G(X(y))} \frac{\partial \varepsilon}{\partial t}-\tilde E 
\frac{\partial j}{\partial y} \qquad .
\end{equation}
while the field equation simplifies to 
\begin{equation}
\varepsilon^{(L)} = \sum_i^{L-1} a_i \varepsilon^{(i)} + K j(1,t) \ .
\label{deml}
\end{equation}
By now introducing the standard Ansatz, $j = j_f \exp(\lambda t)$ and 
$\varepsilon = \varepsilon_f \exp(\lambda t)$,
one obtains the eigenvalue equation \cite{abbott},
\begin{equation}
\left({\rm e}^{\lambda/\tilde E}-1\right) \prod_{k=1}^L (\lambda + \alpha_k) = \frac{g K \lambda}{\tilde E}
 \int_0^1 dy \frac{{\rm e}^{\lambda y/\tilde E}}{G(X(y))}
\label{charac}
\end{equation}
In the case of a constant $G(X(y)) = \sigma$, $L$ eigenvalues correspond to the
zeroes of the following polynomial equation
\begin{equation}
\prod_{k=1}^L (\lambda + \alpha_k) = \frac{g K}{\sigma} \, .
\label{charac2}
\end{equation}
For $g=0$ such solutions are the poles which define the field dynamics, while for
$g=\sigma$, $\lambda=0$ is a solution: this corresponds to the maximal value
of the (positive) coupling strength beyond which the model does no longer support 
stationary states, as the feedback induces an unbounded growth of the spiking rate.
Besides such $L$ solution, the spectrum is composed of an infinite set of purely 
imaginary eigenvalues,
\begin{equation}
\lambda = 2 \pi i n \tilde E = \frac{2 \pi i n}{T_0} \quad n \ne 0 \, .
\end{equation}
The existence of such marginally stable directions reflects the fact that all 
$y^i$ phases experience
the same velocity field, independently of their current value (see Eq.~(\ref{phase})), so that 
no effective interaction is present among the oscillators.
In the limit of small variations of $G(X(y))$, one can develop a perturbative approach.
Here below, we proceed under the more restrictive assumption that the coupling constant $g$ is
itself small: we have checked that this restriction does not change the substance of our 
conclusions, while requiring a simpler algebra.

By assuming that $g$ is small, one can assume that the deviation of $\lambda$ from 
$2\pi in \tilde E$ are small as well and thereby expand the exponential in Eq.~(\ref{charac}). 
Up to first order, we find
\begin{equation}
\lambda_n = 2 \pi i n \tilde E \, \left[ 
1 + \frac{ gK (A_n +i B_n)}{\prod_{k=1}^L (2 \pi i n \tilde E + \alpha_k)}\right]
\label{lamb1}
\end{equation}
where 
\begin{equation}
(A_n +i B_n) = \int_0^1 dy \frac{{\rm e}^{i 2 \pi n y}}{G(X(y))}
\label{an_bn}
\end{equation}
are the Fourier components of the phase-response curve $1/G(X(y))$.

In order to estimate the leading term of the real part of $\lambda$ 
in the large $n$ limit, let us rewrite Eq.~(\ref{lamb1}) as
\begin{equation}
\lambda_n = i \gamma_n + gK\gamma_n \frac{-B_n+iA_n}{ \prod_{k=1}^L (\alpha_k^2 + \gamma_n^2)}
\prod_{k=1}^L (\alpha_k - i \gamma_n)
\label{lamb2}
\end{equation} 
where $\gamma_n = 2 \pi n \tilde E = (2 \pi n)/T_0$. Since $\gamma_n$ is proportional to $n$,
the leading terms in the product at numerator of  Eq.~(\ref{lamb2}) are
$$
\prod_{k=1}^L (\alpha_k - i \gamma_n) \sim (-i)^L \gamma_n^L + S 
(-i)^{L-1} \gamma_n^{L -1}
$$
where $S = \sum_{k=1}^L \alpha_k $. Accordingly, in the case of even $L$,
\begin{equation}
Re\{\lambda_n\} \sim gK (-1)^{L/2} \left[
\frac{S A_n}{\gamma_n^{L}}-
\frac{B_n}{\gamma_n^{L-1}} \right]
\label{lamb_even}
\end{equation}
and for odd $L$,
\begin{equation}
Re\{\lambda_n\} \sim gK (-1)^{(L+3)/2}
\left[
\frac{A_n}{\gamma_n^{L-1}} +
\frac{S B_n}{\gamma_n^{L}} \right] \quad .
\label{lamb_odd}
\end{equation}

For a discontinuous $F(X)$, it can be shown that 
\begin{eqnarray}
A_n &\simeq& \frac{-T_0}{4 \pi^2 n^2} \left[ 
\frac{F^\prime(1)}{G(1)^2}- 
\frac{F^\prime(0)}{G(0)^2} \right]  \, ,
\label{A}
\\
B_n &\simeq& \frac{T_0}{2 \pi n} 
\left[ \frac{F(1)-F(0)}{G(1)G(0)} \right]  \, ,
\label{A_B}
\end{eqnarray}
the procedure to derive the above expressions is briefly
sketched in Appendix \ref{App1}. 

Therefore, for even $L$, the leading term for $n \to \infty$ is
\begin{equation}
Re\{\lambda_n\} = 
\frac{gK T_0^L (-1)^{L/2} \left(F(0)-F(1)\right) }
{ (2 \pi n)^L G(1)G(0)} 
\quad .
\label{exp_even}
\end{equation} 
For even $L$, the stability of the short-wavelength modes (large $n$)
is controlled by the sign of $(F(0) - F(1))$: for even (odd) $L/2$ and
excitatory coupling, i.e. $g>0$, the splay state is stable whenever
$F(1) > F(0)$ ($F(1)<F(0)$). Obviously the stability is
reversed for inhibitory coupling. 

Notice that for $L=0$, i.e. $\delta$-spikes, the eigenvalues do
not decrease with $n$, as previously observed in \cite{zillmer2}. 
This is the only case where all modes exhibit a finite stability 
even in the thermodynamic limit.

For odd $L$, the real part of the eigenvalues is
\begin{eqnarray}
Re\{\lambda_n\} &=&
\frac{gK T_0^L (-1)^{(L+1)/2}}{ (2 \pi n)^{(L+1)}} \times 
\label{exp_odd}
\\
&& \left\{ \frac{F^\prime(1)}{G(1)^2}- \frac{F^\prime(0)}{G(0)^2} -
S T_0 \frac{F(1)-F(0)}{G(1)G(0)} \right\}  \nonumber 
\quad ,
\end{eqnarray} 
in this case the value of $F(X)$ and of its derivative $F^\prime(X)$ at the extrema 
mix up in a nontrivial way.

Finally, as for the scaling behaviour of the leading terms we observe that 
\begin{equation}
{\it Re}\{\lambda_n\} \sim n^{-q}  \, ,  \quad q = 2 \left\lfloor \frac{L+1}{2} \right\rfloor
\label{scaling}
\end{equation}
where $\lfloor\cdot\rfloor$ stays for the integer part of the number. Therefore
the scaling of the short-wavelength modes for discontinuous $F(X)$ is dictated 
by the post-synaptic pulse profile.

For a continuous but non differentiable $F(X)$, (i.e. $F^\prime(1) \ne F^\prime(0)$),
if $L$ is even, it is necessary to go two orders beyond in the estimate of the Fourier coefficients
(see Appendix \ref{App1}). As a result, the eigenvalues scale 
as  
\begin{equation}
Re\{\lambda_n\} \propto n^{-(L+2)}
\, .
\label{exp_even_sca}
\end{equation} 
For odd $L$, it is instead sufficient to assume $F(0)=F(1)$ in Eq.~(\ref{exp_odd}). 

Altogether, we have seen that the non-smoothness of both the post-synaptic pulse and of the velocity field
(or, equivalently, of the phase response curve) play a crucial role in determining the degree
of stability of the splay state. The smoother are such functions and the slower
short-wavelength perturbations decay, although the changes occur in steps 
which depend on the parity of the order of the discontinuity (at least for the pulse structure).
Moreover, the overall stability of the spectral components depends in a complicate way
on the sign of the discontinuity itself.

\section{Theory (finite $N$)} 
\label{sec:3}

\subsection{The splay state}
\label{sec:3:A}

The stability for finite $N$ can be investigated by linearizing 
Eqs.~(\ref{event_driven_map_E},\ref{event_driven_map_P},\ref{generalized_event_driven_map}).
A thorough analysis has been developed in \cite{olmi_math}; here we limit ourselves
to review the key ideas as a guide for the numerical analysis. 

We start by introducing the vector $W = (\{x^j\},\epsilon,p)$ ($j=1,N-1$), whose 
components represent the infinitesimal perturbations of the solution $\{X^j\}$, $E$, $P$.
The Floquet spectrum can be determined by constructing the matrix $\bf A$ which maps 
the initial vector $W(0)$ into $W(\mathcal{T})$,
\begin{equation}
W(\mathcal{T}) = {\bf A} W(0)
\end{equation}
where $\mathcal{T}$ corresponds to the time separation between two consecutive spikes.
This is done in two steps, the first of which corresponds to evolving the components
of a Cartesian basis according to the equations obtained from the linearization of
Eqs.~(\ref{sistema},\ref{sistema2}) (in the comoving reference frame),
\begin{eqnarray}
\nonumber
&& \dot{x}^{j} =\frac{d F}{d x_{j+1}} x^{j+1} + g \epsilon, \quad j=2,\dots,N
\quad \dot{x}^{N} \equiv 0
\\ \label{differenziale}
&& \dot{\epsilon}= p - \alpha \epsilon, \qquad
\dot{p}= -\alpha p \enskip .
\end{eqnarray}
The second step consists in accounting for the spike emission, which amounts to 
to add the vector
\begin{equation}
\label{spazio_tang}
U = [\{\dot{X}^j(\mathcal{T})\}, \dot{E}(\mathcal{T}), \dot{P}(\mathcal{T})] \tau
\end{equation}
where $\tau$ is obtained from the linearization of the threshold
condition (\ref{tauu}), 
\begin{equation}
\label{deltatau}
 \tau = -\left(\frac{\partial X^1}{\partial E} \epsilon + 
\frac{\partial X^1}{\partial P} p \right)\frac{1}{\dot{X}^1} 
\end{equation}
The diagonalization of the resulting matrix $\bf A$, gives $N+1$ Floquet eigenvalues $\mu_k$,
which we express as
\begin{equation}
 \mu_k= e^{i\phi_k}e^{T_0(\lambda_k + i\omega_k)/N},
\end{equation}
where $\phi_k= \frac{2\pi k}{N}$, $k=1,\ldots,N-1$ and $\phi_N=\phi_{N-1}=0$, while
$\lambda_k$ and $\omega_k$ are the real and imaginary parts of the Floquet exponents.
The variable $\phi_k$ plays the role of the wavenumber $k$ in the linear stability analysis 
of spatially extended systems.

Previous studies \cite{olmi_math} have shown that the spectrum can be decomposed 
into two components: (i) $k\sim \mathcal{O}(1)$; (ii) $k/N \sim \mathcal{O}(1)$. 
The former one is the LW component and can be directly obtained in the thermodynamic limit 
(see the previous section). For $L=2$ and $\alpha_1 = \alpha_2$ (i.e. for $\alpha$ pulses), 
it has been found that the results reported in in \cite{abbott} match does obtained for
$1\ll k \ll N$ in \cite{olmi_math}. The latter one corresponds to the SW component: it depends
on the system size and cannot, indeed, be derived from the mean field approach discussed in the 
previous section. In the next section, we
illustrate some examples that go beyond the analytic studies carried out in 
\cite{olmi_math}.

\subsection{The synchronized state} 

In this section we address the problem of measuring the stability of the fully synchronized state
for a generic oscillator dynamics $F(x)$. The task is non trivial, because of the resetting
mechanism, which acts simultaneously on all neurons. On the one side, we extend the results
obtained in \cite{goel} which are restricted to a continuous PRC, on the other side we extend the results of
Mirollo and Strogatz~\cite{mirollo} which refer to excitatory coupling and $\delta$ pulses. 
In order to make the analysis easier to understand
we start considering $\alpha$-pulses. Other cases are discussed afterwards.

The starting point amounts to writing the event driven map in a comoving frame, 
\begin{eqnarray}
 \label{event_driven_map:v}
&& X^{j}_{n+1} = \mathcal{F} \left( X_n^{j+1}, E_n, P_n, \mathcal{T}_n \right) 
\\ \label{event_driven_map:E}
&& E_{n+1}=E_n e^{-\alpha \mathcal{T}_n}+ P_n \mathcal{T}_n e^{-\alpha \mathcal{T}_n}, 
\\ \label{event_driven_map:P}
&& P_{n+1}=P_n e^{-\alpha \mathcal{T}_n} +\frac{\alpha^2}{N}  \, ,
\end{eqnarray}
where the function $\mathcal{F}$ is obtained by formally integrating the equations of motion
over the time interval $\mathcal{T}_n$. Notice that the field dynamics has been, instead, 
explicitly obtained from the exact integration of the equations of motion 
(compare with Eqs.~(\ref{eq:E},\ref{sistema2})). 
The interspike time interval $\mathcal{T}_n$ is finally determined by solving the implicit
equation
\begin{equation}
\label{threshold_condition}
\mathcal{F}( X^1_n, E_n, P_n, \mathcal{T}_n) = 1 . 
\end{equation}
In order to determine the stability of the synchronized state, it is necessary to 
assume that the neurons have an infinitesimally different membrane potentials,
even though they coincide with one another. As a result, the full period must be broken 
into $N$ steps. In the first one, of length $T$, all neurons start in $X=0$ and arrive at 1, but 
only the ``first" reaches the threshold; in the following $N-1$ steps, of 0-length, 
one neuron after the other passes the threshold and it is accordingly reset in 0.

With this scheme in mind we proceed to linearize the equations, writing the evolution
equations for the infinitesimal perturbations $x^j_n$, $\epsilon_n$, $p_n$, and $\tau_n$ 
around the synchronous solution. From Eq.~(\ref{event_driven_map:v}-\ref{event_driven_map:P})
we obtain,
\begin{eqnarray}
x^{j}_{n+1}&=&\mathcal{F}_X(j+1) x^{j+1}_n + \mathcal{F}_E(j+1) \epsilon_n + 
\nonumber \\
&& \mathcal{F}_P(j+1) p_n + \mathcal{F}_{\mathcal{T}}(j+1) \tau_n   \quad 1 \le j < N \label{eq:solx} \\
\epsilon_{n+1}&=& e^{-\alpha\mathcal{T}} \epsilon_n + \mathcal{T} e^{-\alpha\mathcal{T}} p_n -\nonumber \\
&& \left( \alpha \tilde E - P_n e^{-\alpha\mathcal{T}}\right)\tau_n \label{eq:sole} \\ 
p_{n+1}&=& e^{-\alpha\mathcal{T}} p_n -\alpha P_n e^{-\alpha\mathcal{T}}\tau_n
\label{eq:solp} \, .
\end{eqnarray}
with the boundary condition $x^N_{n+1}=0$ (due to the reset mechanism) and
where the subscripts $X$, $E$, $P$, and $\mathcal{T}$ denote a partial derivative with respect to the
given variable. Moreover, the dependence on $j+1$ is a shorthand notation to remind that
the various derivatives depend on the membrane potential of the $(j+1)$st neuron. Finally,
we have left the $n$-dependence in the variable $P$ as it changes (in $\alpha^2/N$ steps,
while the neurons cross the threshold), while $\tilde E$ refers to the field amplitude,
which, instead, stays constant.

The above equations must be complemented by the following condition : 
\begin{equation}
\label{deltataugenerale}
 \tau_n = - \mathcal{T}_X x^{1}_n + \mathcal{T}_E \epsilon_n + \mathcal{T}_P p_n 
\, ,
\end{equation}
where $\mathcal{T}_Z = \mathcal{F}_Z(1)/\mathcal{F}_\mathcal{T}(1)$ ($Z=X$, $E$, $P$).
Eq. (\ref{deltataugenerale}) is obtained by differentiating Eq.~(\ref{threshold_condition})
which defines the period of the splay state.

We now proceed to build the Jacobian for each of the $N$ steps, starting from the first one.
In order not to overload the notations, from now on, the time index $n$ corresponds to the
step of the procedure.
It is convenient to order all the variables, starting from $x^j$ ($j=1,N-1$), and then
including $\epsilon$ and $p$, into a single vector, so that the evolution is described
by an $(N+1)\times(N+1)$ matrix with the following structure, 
\begin{equation}
\mathcal{N}(n)=
\left( 
\begin{array}{cc}
 {\Gamma}(n) & \textbf{0} \\
 {\Psi}(n) & {\Omega}(n)  \\
\end{array}
\right), 
\end{equation}
where $\textbf{0}$ is an $(N-1)\times2$ null matrix;
$\Gamma(n)$ is a quadratic $(N-1)\times(N-1)$ matrix, whose only non-zero elements are those in the
first column and along the supradiagonal; $\Psi(n)$ is a $2\times(N-1)$ matrix 
whose elements are all zero except for the first column; finally $\Omega(n)$ is a
$2\times 2$ matrix.

Since in the first step all neurons start from the same position $X=0$, 
one can drop the $j$ dependence in $\mathcal{F}$. With the help of 
Eqs.~(\ref{deltataugenerale},\ref{eq:solx}) 
\begin{eqnarray}
 &&\Gamma(1)_{j,1} = -\mathcal{F}_X \nonumber \\
 &&\Gamma(1)_{j,j+1} = \mathcal{F}_X 
\end{eqnarray}
Moreover, with the help of Eqs.~(\ref{eq:sole},\ref{eq:solp},\ref{deltataugenerale})
\begin{eqnarray}
\Psi(1)_{11} &=& - \left(\alpha \tilde E - \tilde P e^{-\alpha T} \right)\mathcal{T}_{X} \nonumber \\
\Psi(1)_{12} &=&  - \alpha P e^{-\alpha T}\mathcal{T}_{X} ,
\end{eqnarray}
where we have also made use that $P_1=\tilde P$.
Finally, 
\begin{eqnarray}
\label{abbreviazioni}
&& \Omega(1)_{11} = e^{-\alpha T}-\left(\alpha \tilde E - \tilde P e^{-\alpha T} \right)\mathcal{T}_E,
\nonumber \\
&& \Omega(1)_{12} = T e^{-\alpha T}-\left(\alpha\tilde E -\tilde P e^{-\alpha T} \right)\mathcal{T}_P,
\nonumber \\
&& \Omega(1)_{21} = - \alpha \tilde P e^{-\alpha T}\mathcal{T}_E, \\
&& \Omega(1)_{22} = e^{-\alpha T} - \alpha \tilde P e^{-\alpha T}\mathcal{T}_P,
\nonumber
\end{eqnarray}
In the next steps, $\mathcal{T}_n$ vanishes, so that
$\mathcal{F}_E=\mathcal{F}_P=0$, while $\mathcal{F}_X=1$ and 
$\mathcal{F}_\mathcal{T}(1) = F(1)+g\tilde E:= V^1$. Moreover, $\mathcal{F}_\mathcal{T}(j)$
depends on whether the $j$th neuron has passed the threshold or not. 
In the former case $\mathcal{F}_\mathcal{T}(j+1)=F(0)+g\tilde E:=V_0$, 
otherwise $\mathcal{F}_\mathcal{T}(j+1)=V^1$. As a result,
\begin{eqnarray}
\label{altri_passi}
 &&\Gamma(n)_{j,1} = -V^j/V^1 \nonumber \\
 &&\Gamma(n)_{j,j+1} = 1 
\end{eqnarray}
where $V^j=V^0$ if $j<n$ and $V^j=V^1$, otherwise.
At the same time, from the equations for the field variables, we find that
\begin{eqnarray}
\nonumber
&& \Psi(n)_{11}=\frac{\alpha \tilde E- (\tilde P + (n-1)\frac{\alpha^2}{N})}{V^1} \\
&& \Psi(n)_{12}=\frac{\alpha (\tilde P + (n-1)\frac{\alpha^2}{N})}{V^1}, 
\end{eqnarray}
while $\Omega(n)$ reduces to the identity matrix.

From the multiplication of all matrices, we find that the structure is preserved, namely
\begin{equation}
\mathcal{N}_{N}\cdots\mathcal{N}_2\mathcal{N}_1 =
\left( 
\begin{array}{cc}
 \Lambda & \textbf{0} \\
 \bar{\Psi} & \Omega(1)
\end{array}
\right), 
\end{equation}
where $\Lambda$ is a diagonal matrix, with 
\begin{equation}
\Lambda_{jj} = \mathcal{F}_X \frac{V^0}{V^1}
= \frac{F(0) +g \tilde E}{F(1) +g \tilde E} \exp \left[\int_0^{T} dt F^\prime(X(t))\right]
\label{eigen_synch}
\end{equation}
which therefore measures the stability of the orbit (together with the eigenvalues of
$\Omega$ which are associated to the pulse structure). 
In practice, $\mathcal{F}_X$ corresponds to the expansion rate from $X=0$ to $X=1$ 
under the action of the mean field $E$ and we recover a standard result in 
globally coupled identical oscillators: the spectrum is degenerate, all eigenvalues
being equal and independent of the network size. The result is, however, not obvious in this
context, due to the care that is needed in taking into account the various discontinuities.
We have separately verified that the same conclusion holds for exponential spikes.

The stability of the synchronized state can be also addressed by determining the evaporation 
exponent $\Lambda_e$ \cite{vvres,evap}, which measures the stability of a probe neuron subject 
to the mean field generated by the synchronous neurons with no feedback towards them. 
By implementing this approach for a negative perturbation, van Vreeswijk found that $\Lambda_e$
is equal to $\Lambda_{jj}$ (for $\alpha$-functions). By further assuming that $F^\prime <0$,
he was able to prove that the synchronized state is stable for inhibitory coupling and 
sufficiently small $\alpha$-values.
The situation is more delicate for exponential pulse-shapes. As shown in~\cite{divolo}, 
$\Lambda_e >0$ ($\Lambda_e <0$) depending whether the perturbation is positive (negative).
In this case, the Floquet exponent reported in Eq.~(\ref{eigen_synch}) coincides with 
the evaporation exponent estimated for negative perturbations.
In Appendix~\ref{App2} we show that the difference between the left and right stability 
is to be attributed to the discontinuous shape of the pulse: no anomaly is expected
for $\alpha$ pulses.

\section{Numerical analysis}
\label{sec:4}

The theoretical approaches discussed in the previous sections allow determining: (i) the SW 
components of the Floquet spectrum for discontinuous velocity fields;
(ii) the leading LW exponents directly in the thermodynamic limit for generic velocity
fields and pulse shapes, in the weak coupling limit.
It would be possible to extend the finite $N$ results to other setups, but we do not think that the
effort is worth, given the huge amount of technicalities. We thus prefer to illustrate the expected
behaviour with the help of some simulations which, incidentally, cover a wider range than possibly
accessible to the analytics.

More precisely, in this and the following section we study the models listed in Table~\ref{tab1}
in a standard set up (splay states) and under the effect of periodic external perturbations.

\begin{table}
\begin{tabular}{c|c}
\hline
\hline
Velocity Field  &  Analiticity  \\
\hline 
$F_{0}(X)=a-X$ & Discontinuous\\
$F_1(X)=a-X(X-0.7)$  &  Discontinuous \\
$F_2(X)=a-0.25\sin(\pi X)$ &  ${\cal C}^{(0)}$ \\
$F_3(X)=a+X(X-1)$ &    ${\cal C}^{(0)}$         \\
$F_4(X)=a - 0.25 \sin(\pi X)\cos^2(\pi X)$ & ${\cal C}^{(0)}$   \\
$F_5(X)=a - 0.25 \sin(2\pi X)\cos^{2}(2\pi X)$ & ${\cal C}^{(\infty)}$ \\
$F_6(X)=a - 0.25 \sin(2\pi X)e^{\cos(2\pi X)}$ & ${\cal C}^{(\infty)}$ \\
$F_7(X)=a - 1 + e^{2\sin(2\pi X)}$  & ${\cal C}^{(\infty)}$ \\
\hline\hline
\end{tabular}
\label{tab1}       
\end{table}

\subsection{Finite pulse width}

Here, we discuss the stability of the splay state for different degrees of
smoothness of the velocity field at the borders of the unit interval
for post-synaptic pulses of $\alpha$-function type.

\begin{figure}[h]
\begin{center}
\includegraphics*[width=8.5cm]{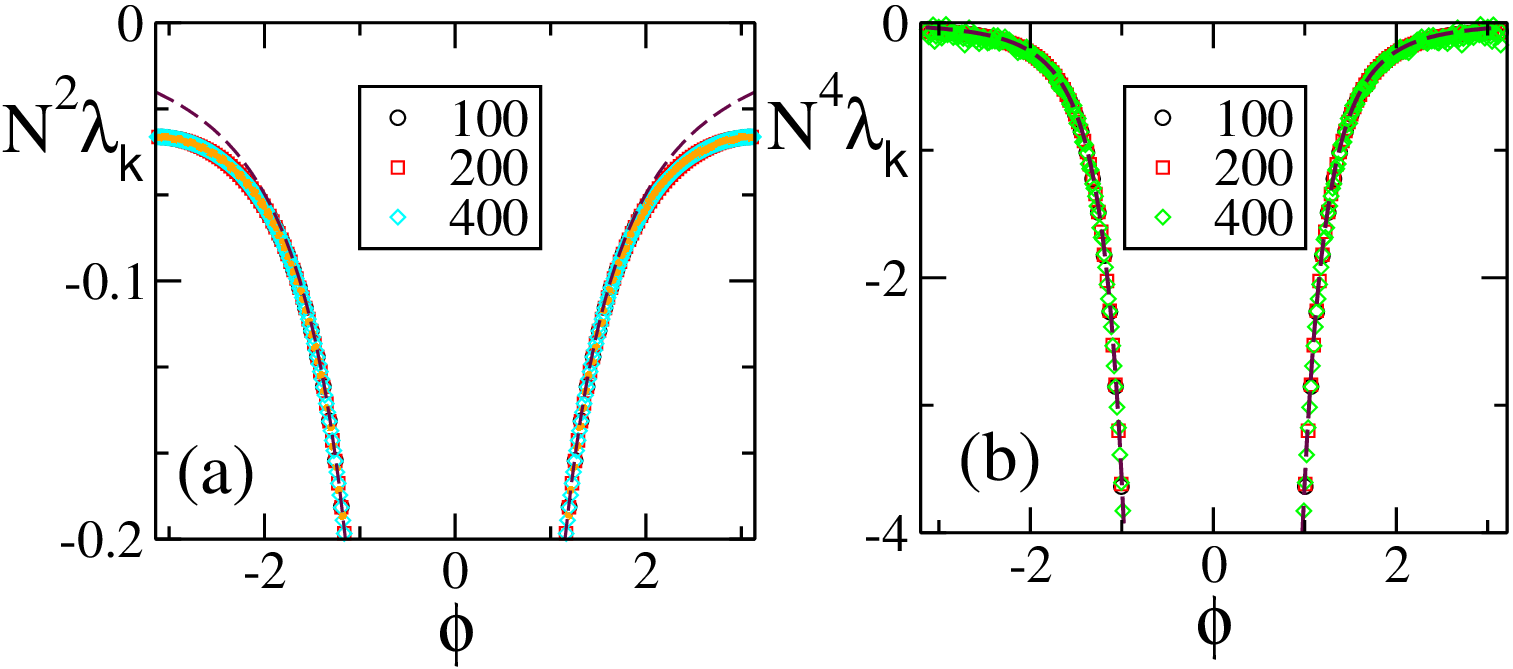}
\end{center}
\caption{Floquet spectra for $\alpha$-pulses for
(a) a discontinuous field $F_1(X)$ and (b) a continuous field $F_4(X)$.
The orange dotted line in panel (a) represents the theorical curve estimated by using 
Eq. (7) in \cite{olmi_math}, while the dashed maroon curve represents the theorical curve
estimated by using Eq.~(\ref{exp_even}) in Sec. 3. In panel (b) the dashed maroon curve
is calculated by using Eq.~(\ref{exp_even_con}). 
All data refer to $a=1.3$ and $\alpha=3$.}
\label{fig0} 
\end{figure}

We start from discontinuous velocity fields. They have been the subject of an analytic 
study which proved that the SW component scales as $1/N^2$~\cite{olmi_math}.
The data reported in Fig.~\ref{fig0}(a) for $F_1(X)$ confirms the expected scaling:
the agreement with the theoretical curve derived in \cite{olmi_math}
is impressive over the entire spectral range, while the mean field Eq.~(\ref{exp_even}) 
gives a very good estimation of the spectrum except for the shortest wavelengths, where it overestimates
the numerical data. The mean field approximation turns out to be more accurate for 
continuous velocity fields (with a discontinuity of the first derivative at the borders of 
the definition interval). Indeed the agreement between the theoretical expression Eq.~(\ref{exp_even_con})
and the numerical data is very good for the entire range (see Fig.~\ref{fig0}(b) which refers to $F_4(X)$).

\begin{figure}[h]
\begin{center}
\includegraphics*[width=8.5cm]{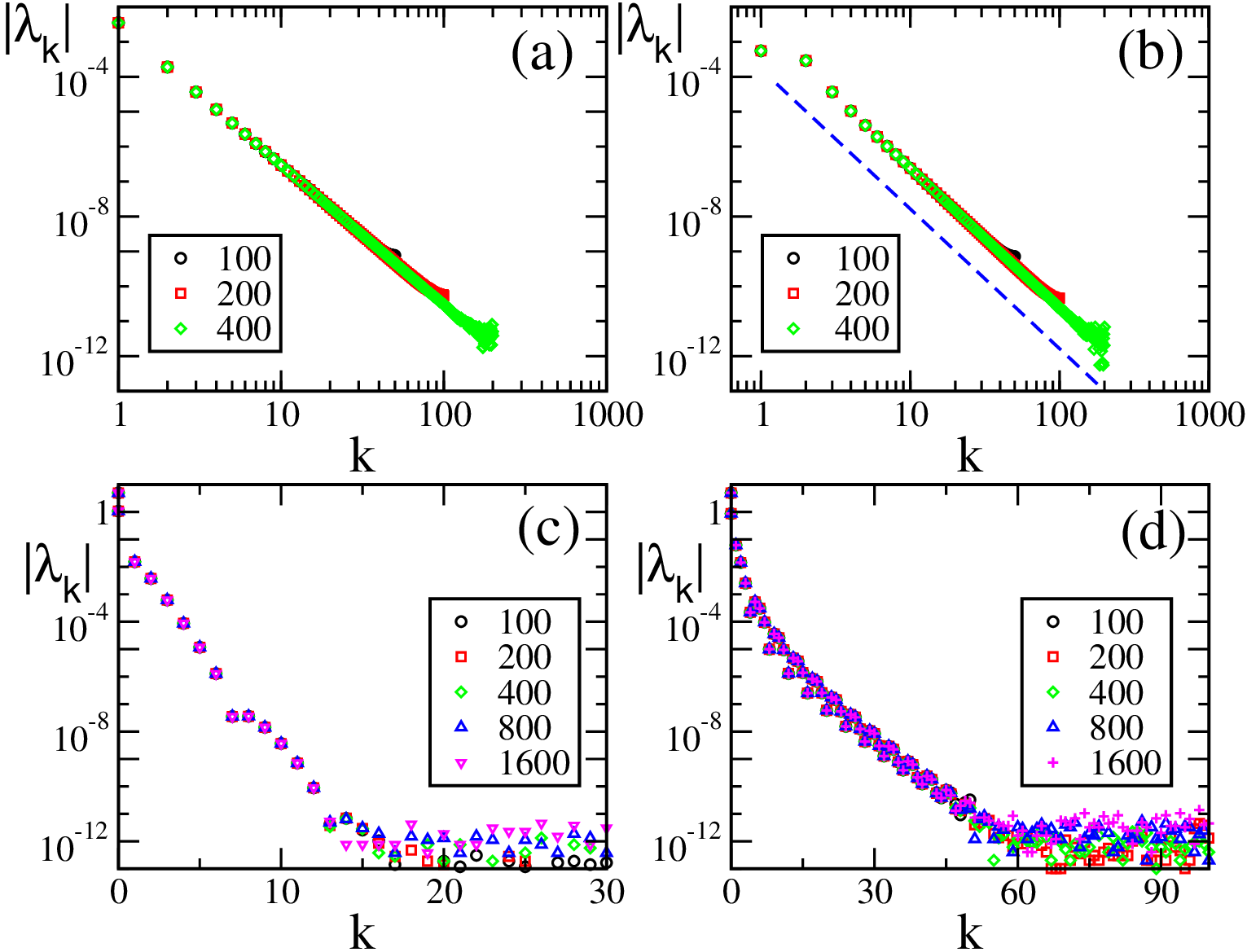}
\end{center}
\caption{Floquet spectra for $\alpha$-pulses for 
two continuous sinusoidal fields, namely $F_2(X)$ (a) and $F_4(X)$ (b); 
and two analytic fields, namely $F_6(X)$ (c) and $F_7(X)$ (d).
The dashed blue line in (b) indicates a scaling $1/k^4$. All data refer to $a=1.3$ and $\alpha=3$.}
\label{fig1} 
\end{figure}

The numerical Floquet spectra for fields that are $\mathcal{C}^{(0)}$, but not $\mathcal{C}^{(1)}$ 
($F(0)=F(1)$, $F'(0)\ne F'(1)$), are reported in Fig.~\ref{fig1} (the curves in panel (a) and (b) refer to 
$F_2(X)$ and $F_4$, respectively). 
For these velocity fields, we have also verified that the spectra scale as $1/N^4$, 
confirming the observation reported in \cite{calamai} for a different velocity field with the same 
analyticity properties. 
The data displayed in Fig.~\ref{fig1} (a,b) refer to the LW components: they indeed confirm 
to be independent of the system size and scale as $1/k^4$ (see the dashed line) as predicted by 
the perturbative theory discussed in section \ref{sec:2}.

The spectra reported in the other two panels  refer to analytic velocity fields:
in all cases the initial part of the Floquet spectra is again independent of $N$ and scales 
approximately exponentially with $k$, confirming that the scaling behavior of the exponents is
related to the analyticity of the velocity field. The fluctuating background with approximate
height $10^{-12}$ is just a consequence of the finite numerical accuracy. This is the reason
why we did not dare to estimate the SW components that would be exceedingly small.

\subsection{Vanishing pulse-width} 

Here, we analyse the intermediate case between finite pulse-width and $\delta$-like impulses.
Similarly to what done in~\cite{zillmer2} for the LIF, we consider $\alpha$ pulses, 
where $\alpha = \beta N$, with $\beta$ independent of $N$.

\begin{figure}[h]
\begin{center}
\includegraphics*[width=8.5cm]{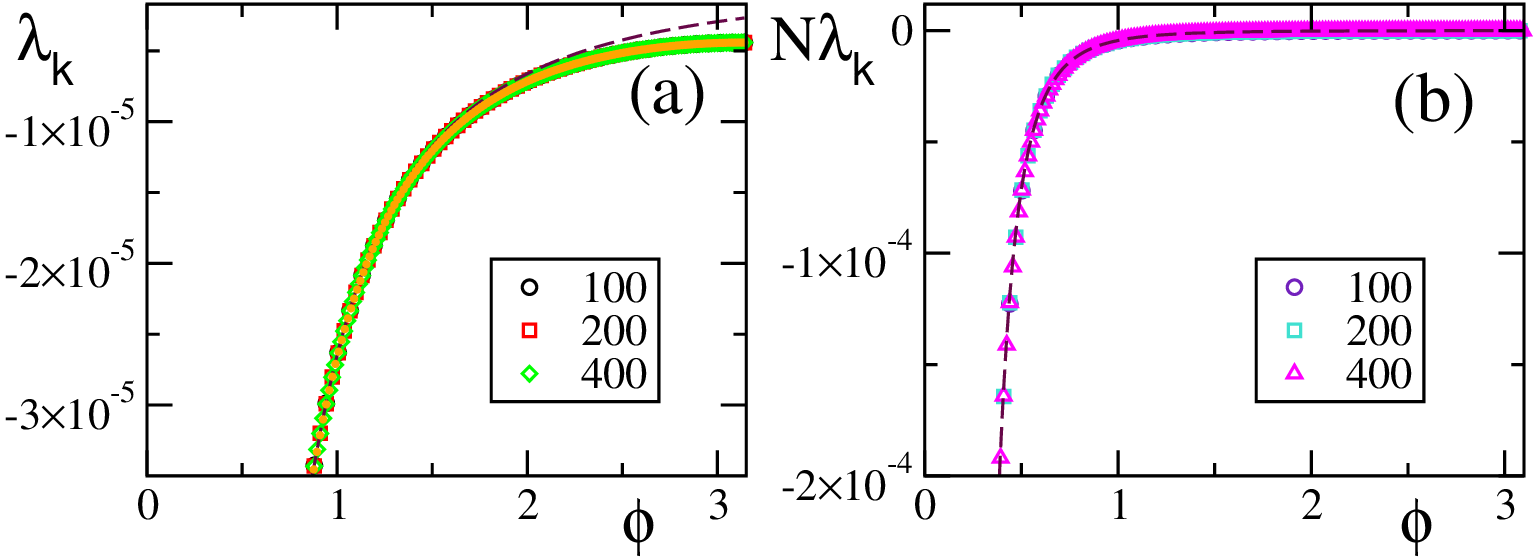}
\end{center}
\caption{Floquet spectra for $\beta$-pulses
with a discontinuous field [$F_1(X)$] (a) and a ${\cal C}^{(0)}$ field [$F_2(X)$] (b).
The orange dotted line in panel (a) represents the theorical curve estimated by using 
Eq.~(7) in \cite{olmi_math}.
The dashed line in panel (a) (resp. (b)) represents the theoretical curve computed by using 
Eq.~(\ref{exp_even}) (resp. Eq.~(\ref{exp_even_con})) for $\beta$-pulses.
The data refer to $a=1.3$ and $\beta=0.03$.}.
\label{fig2} 
\end{figure}

In Fig.~\ref{fig2}(a) we report the spectra for a discontinuous velocity field,
$F_1(x)$. In this case the Floquet spectra remain finite,
so that the corresponding states remain robustly stable even in the thermodynamic limit.
Also in this case the agreement with the theoretical expression reported in Eq. (7) in
\cite{olmi_math} is extremely good, while Eq.~(\ref{exp_even}) overestimates the spectra
for large phases. 
The field considered in panel (b) ($F_2(X)$) is $\mathcal{C}^{(0)}$ but not $\mathcal{C}^{(1)}$. 
In this case, the Floquet spectra scale as $1/N$: this scaling is predicted by the analysis
reported in Sect. \ref{sec:2} and the whole spectrum is very well reproduced
by Eq.~(\ref{exp_even_con}).

\begin{figure}[h]
\begin{center}
\includegraphics*[width=6cm]{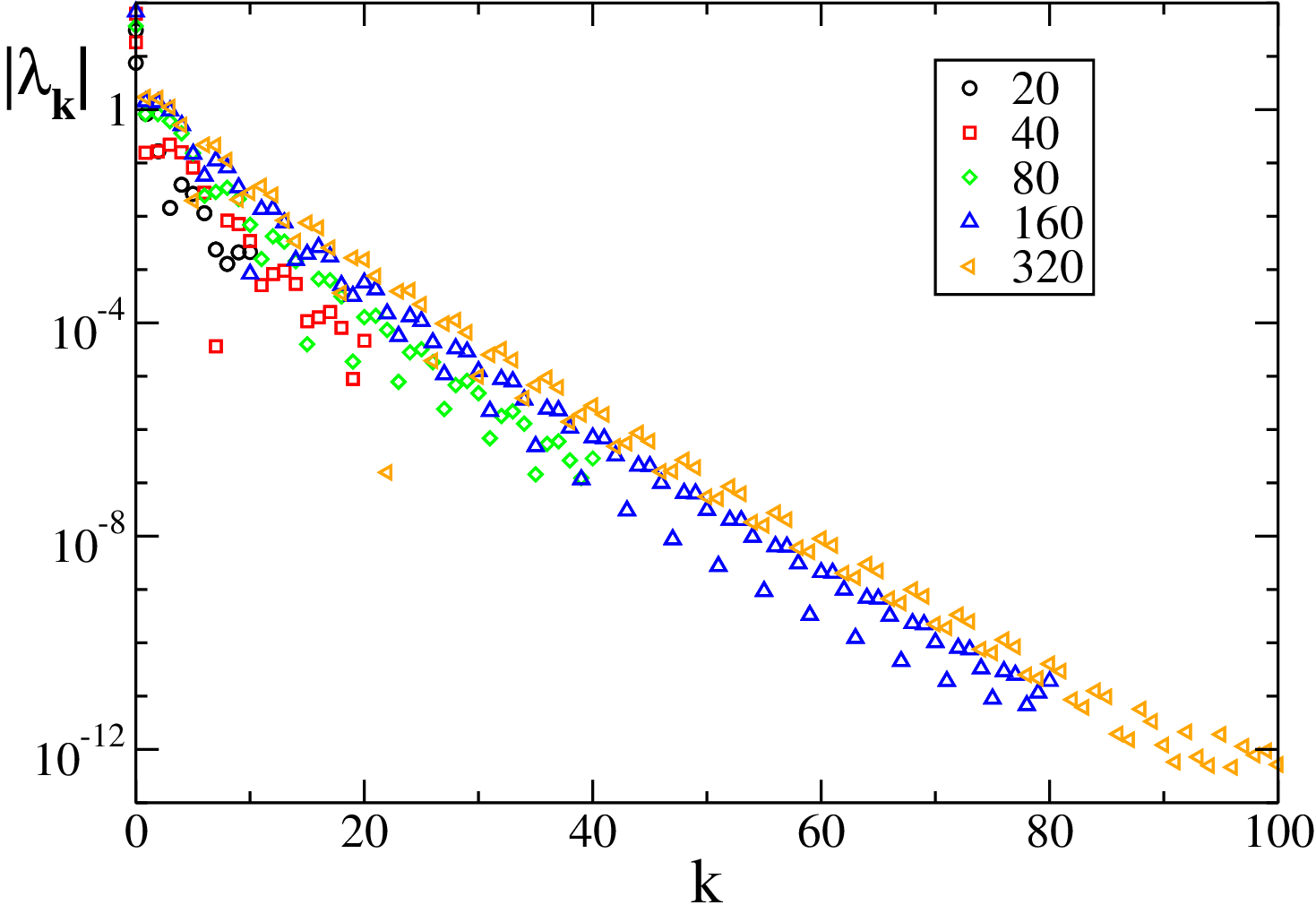}
\end{center}
\caption{Floquet spectra for $\beta$-pulses for 
the analytic field $F_7(X)$. The data refer to $a=1.3$ and $\beta=0.03$.}
\label{fig3} 
\end{figure}

Finally, we have studied an analytic field, namely $F_7(X)$. In this case
the Floquet spectra appear to scale exponentially  to zero with
the wavevector $k$, similarly to what observed for the finite pulse
width, as shown in  Fig.~\ref{fig3}.

\subsection{$\delta$ pulses} 

Finally we considered the case of $\delta$-pulses: whenever the potential $X^j$ reaches the threshold
value, it is reset to zero and a spike is sent to and 
\textit{instantaneously} received by all neurons. We studied just two cases: (i) 
the analytic field $F_7(X)$; (ii) a leaky integrate-and fire neuron model with $F_0(X)$.
The results, obtained for inhibitory coupling (since the splay state is known to be stable 
only in such a case \cite{zillmer,vvres}) are consistent with the expectation for the $\beta$ model.

In particular we found, in the analytic case (i), that the Floquet spectra decay exponentially to zero. 
The exponential scaling is not altered if a phase shift $\zeta$ is introduced in the velocity field
(i.e. for $F(X)=a-1 + e^{2\sin(2\pi X + \zeta)}$).
In the case of the LIF model ($F_0$), we already know that the Lyapunov spectrum tends, in the 
$\delta$-pulse limit, to~\cite{zillmer2}
\begin{equation}
 \lim_{\beta\rightarrow\infty} \lambda_\pi= -1 +\frac{1}{T_0}\ln(\frac{a}{a-1}).
\end{equation}
This result is confirmed by our simulations which also reveal that
the splay state is stable even for small, excitatory coupling values,
extending previous results limited to inhibitory coupling~\cite{zillmer}.

\section{Periodic forcing}
\label{sec:5}

In this Section we numerically investigate the scaling behavior of the Floquet spectrum in the presence
of a periodic forcing, to test the validity of the previous analysis in a more general context. 
We have restricted our studies to splay-state-like regimes, 
where it is important to predict the behavior of the many almost marginally
stable directions. Moreover, we have considered only the smooth $\alpha$-pulses. In this case,
the dynamical equations read 
\begin{eqnarray}
\nonumber
&&\dot{X}^j = F(X^j) + gE + A\cos(\varphi), \qquad j=1,\ldots,N,
\\ \label{forzante}
&&\dot{E} = P - \alpha E,
\\ \nonumber
&&\dot{P} = -\alpha P,
\\ \nonumber
&&\dot{\varphi} = \omega \, .
\end{eqnarray}
They have been written in an autonomous form, since it is more convenient
to perform the Poincar\'e section according to the spiking times, rather than introducing a
stroboscopic map. The interspike interval is determined by the equation
\begin{equation}
\label{eq:tau}
 \mathcal{T}=\int_{X_{old}}^1 \frac{dX^1}{F(X^1)+gE +A\cos(\varphi)}.
\end{equation}
where $X^1$ is the membrane potential of the first neuron (the closest to threshold), 
and ${X_{old}}$ is its initial value.

We analyzed only those setups where the unperturbed splay state is stable.
More precisely: the two discontinuous fields $F_0(X)$ and $F_1(X)$,
the two ${\cal C}^{(0)}$ fields ($F_2(X)$ and $ F_3(X)$), and the analytic field $ F_7(X)$.
In all cases the external modulation induces a periodic modulation of
the mean field $E$ with a period $T_a=2\pi/\omega$ equal to the period of the modulation. 
At the same time, we have verified that, although the forcing term has zero average
(i.e. it does not change the average input current), the average interspike interval
is slightly self-adjusted and, what is more important, there is no evidence of locking 
between the modulation and the frequency of the single neurons. In other
words, the behavior is similar to the spontaneous partial synchronization
observed in \cite{vvres} (where the modulation is self-generated). 

Because of the unavoidable oscillations of the interspike intervals, 
it is necessary to identify the spike times with great care. In practice we integrate 
Eqs.~(\ref{forzante}) with a fixed time step $\Delta t$, by employing a standard fourth-order 
Runge-Kutta integration scheme. At each time step we check if $X^1>1$, in which case we go one
step back and adopt the H\'enon trick, which amounts to exchanging $t$ and $X^1$ in the role
of independent variable \cite{henon1982}.

The linear stability analysis can be performed by linearizing the system (\ref{forzante}),
to obtain 
\begin{eqnarray}
\nonumber
&&\dot{x}^j = \frac{dF(X^j)}{dX^j} x^j+g \epsilon - A\sin(\varphi)\delta\varphi, \qquad j=1,\ldots,N,
\\ \nonumber
&&\dot{\epsilon} = p - \alpha \epsilon,
\\ \nonumber
&&\dot{p} = -\alpha p,
\\ \nonumber
&&\delta\dot{\varphi} =0;
\end{eqnarray}
and by thereby estimating the corresponding Lyapunov spectrum. 

\begin{figure}[t!]
\includegraphics[width=6cm]{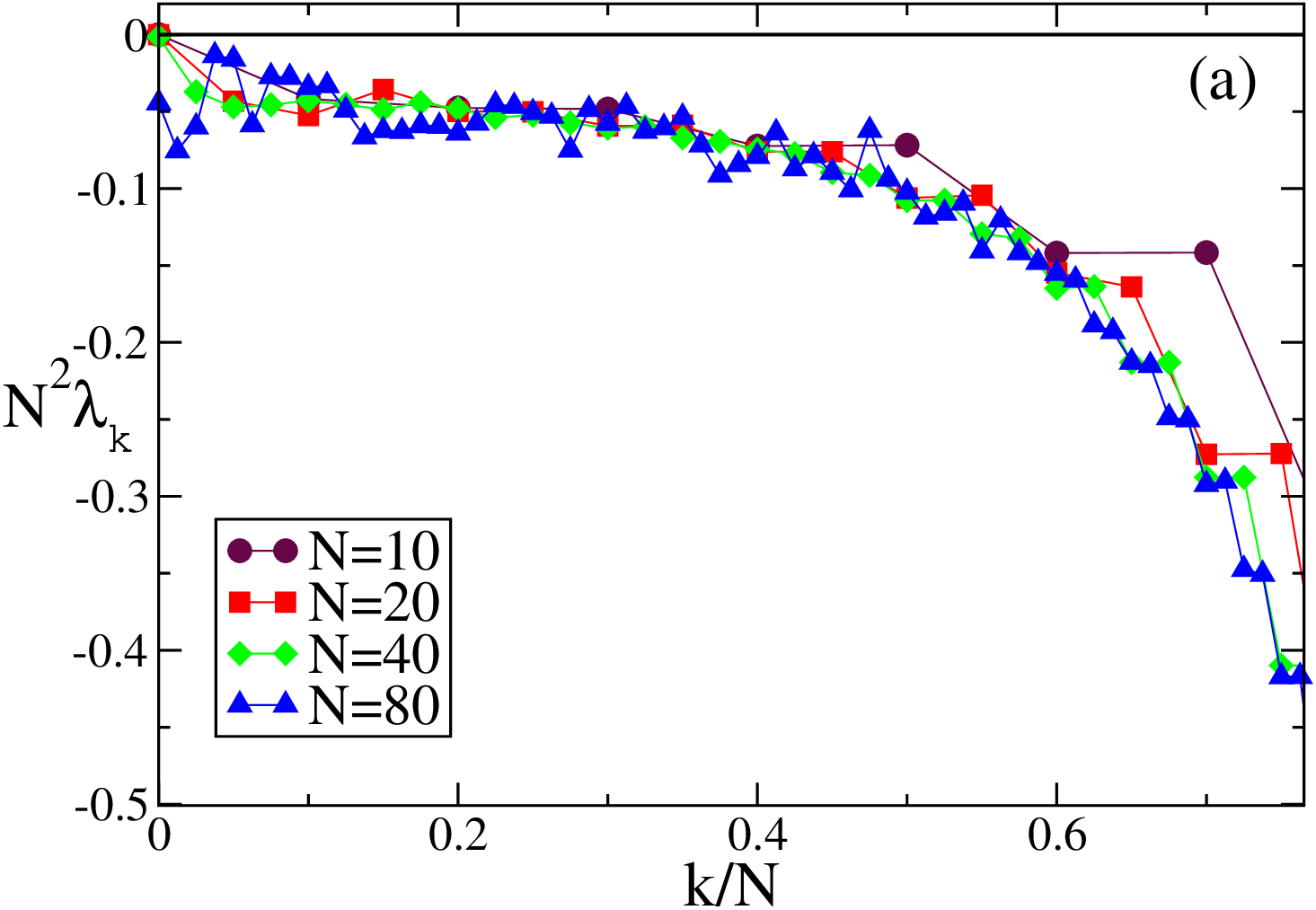}
\includegraphics[width=6cm]{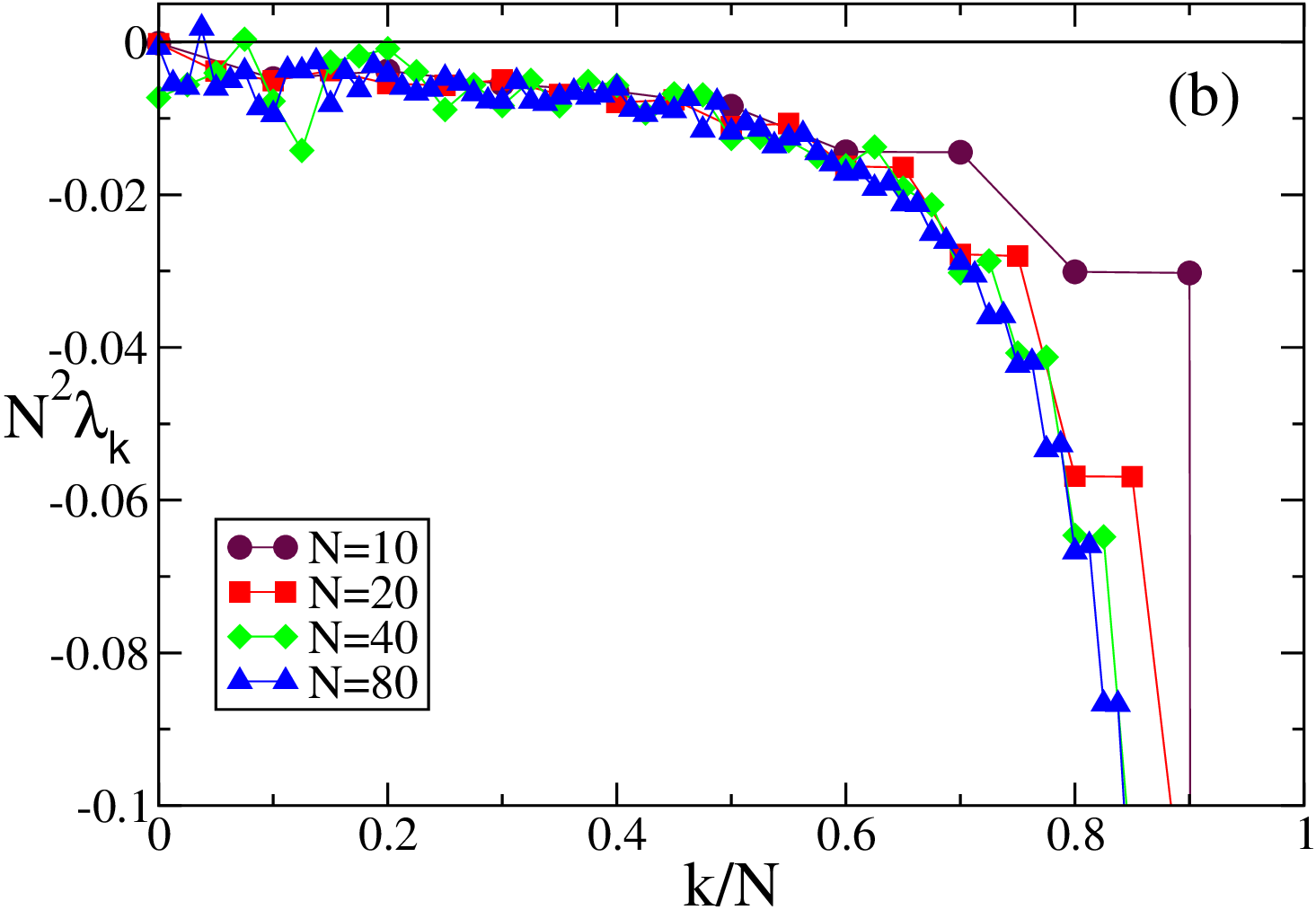}
\caption{Lyapunov spectra for neurons forced by an external periodic signal, 
we observe the scaling $1/N^2$ for the discontinuous velocity fields (a) $F_0(X)$ and (b) $F_1(X)$.
In both cases $A=0.1,T_a=2$. }
\label{forzanteesterna1}
\end{figure}

In the case of $F_0$ and $F_1$, we have always found that the Lyapunov spectrum scales 
as $1/N^2$ as theoretically predicted in the absence of external modulation (see Fig.~\ref{forzanteesterna1}
for one instance of each of the two velocity fields).

\begin{figure}[t!]
\includegraphics[width=6cm]{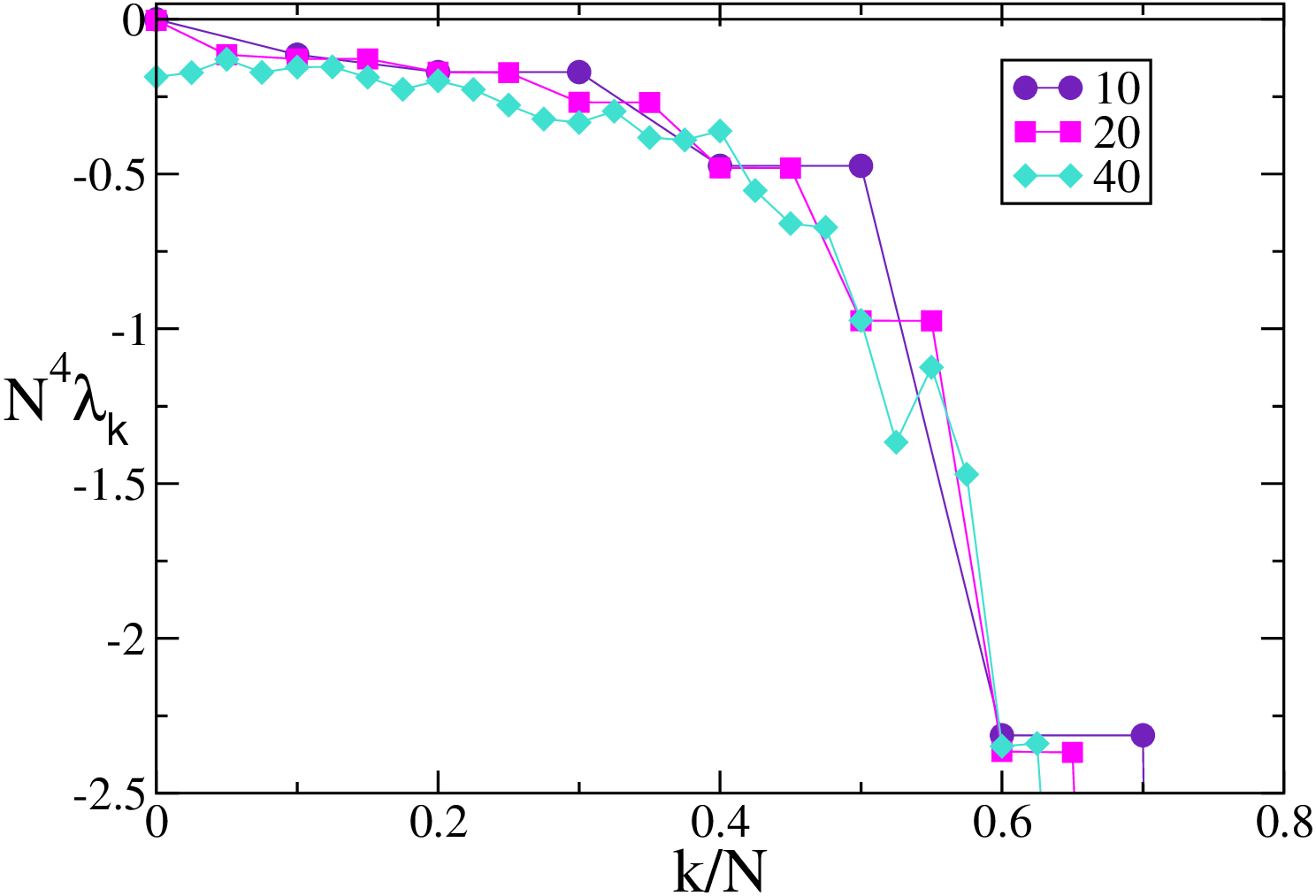}
\caption{Lyapunov spectra for neurons forced by an external periodic signal, 
we observe the scaling $1/N^4$ for the continuous velocity field $F_3(X)$.
The  data refer to $A=0.1,T_a=2$. }
\label{forzanteesterna2}
\end{figure}

A similar agreement is also found for $F_3$, where the Lyapunov spectrum scales as $1/N^4$, exactly as in
the absence of external forcing (see Fig.~\ref{forzanteesterna2}). Analogous results have been obtained
for the other velocity fields (data not shown), which confirm that the validity of the previous
analysis extends to more complex dynamical regimes, as long as the membrane potentials are
smoothly distributed.

\section{Summary and open problems} 
\label{sec:6}
 
In this paper we have discussed the linear stability of both fully synchronized and splay states 
in pulse-coupled networks of identical oscillators. 
By following~\cite{abbott}, we have obtained analytic expressions for the long-wavelength components 
of the Floquet spectra of the splay state for generic velocity fields and post synaptic potential profiles. 
The structure of the spectra depends on the smoothness of both the velocity field and the transmitted pulses. 
The smoother they are and the faster the eigenvalues decrease with the wavelength of the corresponding
eigenvectors. In practice, while splay states arising in LIF neurons with $\delta$-pulses have a finite
degree of (in)stability along all directions, those emerging in analytic velocity fields have many
exponentially small eigenvalues.
Interestingly, the scaling behaviour of the eigenvalues carries over to that of the Lyapunov exponents,
when the network is periodically forced, suggesting that our results have a relevance that goes beyond
the highly symmetric solutions studied in this paper.

Finally, we derived an analytic expression for the Floquet spectra for the fully synchronous state.
In this case the exponents associated to the dynamics of the membrane potentials are all identical,
as it happens for the diffusive coupling, but here the result is less trivial, due to the fact
that one must take into account that arbitrarily close to the solution, the ordering of the neurons may 
be different. Moreover, the value of the (degenerate) Floquet exponent coincides with the evaporation exponent
\cite{vvres,evap} whenever the pulses are sufficiently smooth, while for discontinuous pulses 
(like exponential and $\delta$-spikes) the equivalence is lost (see also \cite{divolo}).

Another important property that has been confirmed by our analysis is the role of the ratio 
$R = N/(T_0\alpha)$ between the width of the single pulse ($1/\alpha$) and the average interspike interval 
of the whole network (${\cal T} = T_0/N$). In fact, it turns out that the asynchronous regimes 
can be strongly stable along all directions only when $R$ remains finite in the thermodynamic
limit (and is possibly small). This includes the idealized case of $\delta$-like
pulses, but also setups where the single pulses are so short that they can be resolved by the
single neurons. Mathematically speaking, this result implies that the thermodynamic limit does not
commute with the limit of a zero pulse-width. It would be interesting to check to what extent 
this property extends to more realistic models. A first confirmation result is contained in \cite{pazo}, 
where the authors find a similar property in a network of Winfree oscillators.

Among possible extensions of our analysis, one should definitely mention the inclusion 
of delay in the pulse transmission. This generalization is far from trivial as it modifies the phase diagram 
of the possible states (see \cite{bar2012} for a recent brief overview of the possible scenarios)
and it complicates noticeably the stability analysis of the synchronized phase. 
An analytic treatment of this latter case is reported in \cite{timme} 
for generic velocity fields and excitatory $\delta$-pulses.

\begin{acknowledgments}
We thank David Angulo Garcia for the help in the use of symbolic algebra
software. AT acknowledges financial support from the European Commission 
through the Marie Curie Initial Training Network 'NETT', project N. 289146,
as well as from the Italian Ministry of Foreign Affairs for
the activity of the Joint Italian-Israeli Laboratory on Neuroscience.
SO and AT thanks the Italian MIUR project CRISIS LAB PNR 2011-2013 for economic support
and the German Collaborative Research Center SFB 910 of the 
Deutsche Forschungsgemeinschaft for the kind hospitality at Physikalisch-Technische 
Bundesanstalt in Berlin during the final write up of this manuscript.
\end{acknowledgments}

\appendix
\section{Fourier Components of the Phase Response Curve}
\label{App1}

In this appendix we briefly outline the way the explicit expression of $A_n$ and $B_n$, 
defined in Eq.~(\ref{an_bn}), can be derived in the large $n$ limit for a velocity field $F(X)$ 
that is either discontinuous, or continuous with discontinuous first derivatives at the
border of the definition interval.

The integration interval $[0,1]$ appearing in Eq. (\ref{an_bn}) is splitted
in $n$ sub-intervals of lenght $1/n$, and the original equation can be rewritten as
\begin{equation}
(A_n +i B_n) = \sum_{k=1}^n \int_{(k-1)/n}^{k/n} dy \frac{{\rm e}^{i 2 \pi n y}}{G(y)} \quad.
\label{an_bn_dis}
\end{equation}
For $n$ sufficiently large we can assume that the variation of $1/G(y)$ is quite limited
within each sub-interval, and we can approximate the function as follows, up to the second order
\begin{eqnarray}
&&\frac{1}{G(y)} = \frac{1}{g+T_0 F(y_0)} \left\{ 1 - \frac{T_0 F^\prime(y_0)}{g+T_0 F(y_0)}(y-y_0)
\right.
\nonumber
\\
&&
+ 
\left.
\left[ \left(\frac{T_0 F^\prime(y_0)}{g+T_0 F(y_0)}\right)^2 - \frac{T_0 F^{''}(y_0)}{2(g+T_0 F(y_0))}\right]
(y -y_0)^2 \right\}
\nonumber
\end{eqnarray}
where $y_0 = (k-1)/n$ is the lower extremum of the $n$th sub-interval.

By inserting these expansions into Eq.~(\ref{an_bn_dis}) and by performimg the integration
over the $n$ sub-intervals, we can determine an approximate expression for $A_n$ and $B_n$.
The estimation of $A_n$ involves integrals containing $\cos(2 \pi n y)$; it is easy to show that
the integral over each sub-interval is zero if the integrand, which multiplies the cosinus term, 
is constant or linear in $y$;
therefore the only non-zero terms are, 
\begin{equation}
\int_{(k-1)/n}^{k/n} dy \cos(2 \pi n y) y^2 = \frac{1}{2 \pi^2 n^3}  \, .
\end{equation}
This allows to rewrite
\begin{eqnarray}
A_n &=& \frac{1}{2 \pi^2 n^2} \sum_{k=1}^n H_2\left(\frac{k-1}{n}\right) \frac{1}{n} 
\nonumber
\\
&=& \frac{1}{2 \pi^2 n^2} \left[ \int_0^1 dx H_2(x)\right] + {\cal O}(\frac{1}{n^3})
\label{aan}
\end{eqnarray}
where 
\begin{equation}
H_2(x) = \left[ \frac{(T_0 F^\prime(x))^2}{(g+T_0 F(x))^3} - \frac{T_0 F^{''}(x)}{2(g+T_0 F(x)^2)}\right]
\, .
\end{equation}
It is easy to verify that $H_2(x)$ admits an exact primitive and therefore to perform the
integral appearing in Eq.~(\ref{aan}) and to arrive at the expression reported in Eq.~(\ref{A}).

The estimation of $B_n$ is more delicate, since now integrals containing $\sin(2 \pi n y)$ are involved.
The only vanishing integrals over the sub-intervals are those with a constant integrand
multiplied by the sinus term and therefore the estimation of $B_n$ reduces to 
\begin{eqnarray}
&& B_n = \sum_{k=1}^n  H_1\left(\frac{k-1}{n}\right) \int_{(k-1)/n}^{k/n} dy \sin(2 \pi n y) y
\nonumber
\\
&+& \sum_{k=1}^n   H_2\left(\frac{k-1}{n}\right) \int_{(k-1)/n}^{k/n} dy \sin(2 \pi n y) 
\left(y^2 -2y \frac{k-1}{n}\right)
\nonumber
\end{eqnarray}
where 
\begin{equation}
H_1(x) = - \frac{T_0 F^\prime(x)}{(g+T_0 F(x))^2}
\qquad ,
\end{equation}
and the non-zero integrals are 
\begin{equation}
\int_{(k-1)/n}^{k/n} dy \sin(2 \pi n y) y = - \frac{1}{2 \pi n^2} 
\qquad ,
\end{equation}
and
\begin{equation}
\int_{(k-1)/n}^{k/n} dy \sin(2 \pi n y) y^2 = \frac{1-2 k}{2 \pi n^3} 
\qquad .
\end{equation}

This allows to rewrite $B_n$ as
\begin{eqnarray}
B_n &=& - \frac{1}{2 \pi n} \sum_{k=1}^n H_1\left(\frac{k-1}{n}\right) \frac{1}{n}
\nonumber
\\
&-& \frac{1}{2 \pi n^2} \sum_{k=1}^n H_2\left(\frac{k-1}{n}\right) \frac{1}{n}
\qquad .
\label{bbn}
\end{eqnarray}
We can then return to a continuous variable by rewriting (\ref{bbn}),
up to the ${\cal O} (1/n^3)$, as 
\begin{eqnarray}
B_n &=& - \frac{1}{2 \pi n} \left[\int_0^1 H_1(x) dx + \frac{H_1(1) - H_1(0)}{2n}\right]
\nonumber
\\
&-& \frac{1}{2 \pi n^2} \int_0^1 H_2(x) dx
\qquad .
\label{bbn1}
\end{eqnarray}
The expression Eq.~(\ref{A_B}) is finally obtained by
noticing that the primitive of $H_2(x)$ is $H_1(x)/2$,
and that
$$
\int_0^1 H_1(x) dx = 
\frac{1}{(g+T_0 F(0))} - \frac{1}{(g+T_0 F(1))} \, .
$$
For continuous velocity fields, $B_n=0$ so that, we can derive from Eq.~(\ref{lamb_even}) 
an exact expression for the real part of the Floquet spectrum in the case of even $L$ 
(for odd $L$ the equivalent expression is given by Eq.~(\ref{exp_odd}))
\begin{equation}
Re\{\lambda_n\} =
\frac{gK S T_0^{L+1} (-1)^{L/2}}{ (2 \pi n)^{(L+2)}} 
\frac{F^\prime(0)-F^\prime(1)}{G(1)^2}  \, .
\label{exp_even_con}
\end{equation} 
A rigorous validation of the above formula would require going one order beyond in the 
$1/n$ expansion of $B_n$, a task that is utterly complicated.
In the specific case of the Quadratic Integrate and Fire neuron (or $\Theta$-neuron) 
$F(X)=a-X(X-1)$, it can be, however, analytically verified that $B_n$ is exactly zero.
Moreover, Eq.~(\ref{exp_even_con}) is in very good agreement with the numerically estimated Floquet spectra
for two other continuous velocity fields, namely $F_4(X)$ and $F_2(X)$ as shown in Fig.~\ref{fig0}
and Fig.~\ref{fig2}, respectively. As a consequence, it is reasonable to conjecture that
Eq.~(\ref{A_B}) is correct up to order ${\cal O} (1/n^4)$.

\section{Evaporation Exponent for the LIF model}
\label{App2}

In this appendix we determine the (left and right) evaporation exponent for a synchronous state
of a network of LIF neurons. This is done by estimating how the potential of a probe neuron, forced by the
mean field generated by the network activity, converges towards the synchronized state.
The stability analysis is performed by following the evolution of a perturbed probe neuron. 
Let us firt consider an initial condition, where the synchronized cluster has just reached 
the threshold $(X_{c} = 1)$, while the probe neuron is lagging behind at a distance $\delta_i$.
Such a distance is equivalent to a delay $t_d$
\begin{equation}
t_d= \frac{\delta_i}{F^{+}(1)} \, ,
\end{equation}
where the subscript ``+" means that the velocity field is estimated just after the pulses have been
emitted. Over the time $t_d$, the potential of the cluster increases from the reset value $0$ to 
\begin{equation}
\delta_c= F^{+}(0) t_d= \frac{F^{+}(0)}{F^{+}(1)}\delta_i \, .
\end{equation}
From now on (in LIF neurons), the distance decreases exponentially, reaching the value
\begin{equation}
 \delta_f=\delta_c e^{-T},
\end{equation}
after a period $T$. As a result,
\begin{equation}
\label{ritardo}
 \frac{\delta_f}{\delta_i}=\frac{F^{+}(0)}{F^{+}(1)}e^{-T}= \frac{a+gE^+}{a-1+gE^+} \, .
\end{equation}
The logarithm of the expansion factor gives the left evaporation exponent
\begin{equation}
 \Lambda_e^{l}=\ln\left( \frac{a+gE^+}{a-1+gE^+}\right)  -T.
\end{equation}

Let us now consider a probe neuron which precedes the synchronized cluster by an amount $\delta_i$. 
After a time $T$ the distance becomes
\begin{equation}
\delta_c = \delta_i e^{-T}
\end{equation}
since no reset event has meanwhile occurred.
Such a distance corresponds to a delay
\begin{equation}
t_d= \frac{\delta_c}{F^{-}(1)} \, ,
\end{equation}
where the subscript ``$-$" means that the velocity has now to be estimated just
before the pulse emission. By proceeding as before one obtains,
\begin{equation}
\label{anticipo}
\frac{\delta_f}{\delta_i}= \frac{F^{-}(0)}{F^{-}(1)}e^{-T} \, .
\end{equation}
so that the right evaporation exponent writes
\begin{equation}
 \Lambda_e^{r}=\ln\left( \frac{a+gE^-}{a-1+gE^-}\right)  -T.
\end{equation}
It is easy to see that the left and right exponents differ if and only if
$E^- \ne E^+$, i.e. if the pulses themselves are not continuous: this is, for
instance, the case of exponential and $\delta$ pulses.


\begin{thebibliography}{}
\bibitem{note1} Upon changing variables, the velocity field
can be made independent of the local variable (as intuitively
expected for a true phase). When this is done, the phase
dependence of the velocity field is moved to the coupling function, 
i.e. to the PRC.
\bibitem{note2} A small effective coupling can arise also when PRC has a very weak dependence
on the phase (see below Sec.~\ref{sec:2}).
\end{thebibliography}


\end{document}